\documentclass[a4paper]{aa}
\usepackage{graphicx}
\usepackage{amssymb}
\usepackage{natbib}
\bibpunct{(}{)}{;}{a}{}{,} 
\usepackage{bm}

\begin{document}
\title{A deeply embedded young protoplanetary disk around L1489~IRS observed by 
	   the Submillimeter Array}
\titlerunning{A deeply embedded young protoplanetary disk around L1489~IRS}
\author{C.~Brinch\inst{1} 
   \and A.~Crapsi\inst{1} 
   \and J.~K.~J{\o}rgensen\inst{2,3}
   \and M.~R.~Hogerheijde\inst{1}
   \and T.~Hill\inst{1}}
\date{}
\institute{Leiden Observatory, P.O. Box 9513, 2300 RA Leiden, The Netherlands\\
     \email{brinch@strw.leidenuniv.nl}
     \and Harvard-Smithsonian Center for Astrophysics, 60 Garden Street, Mail
	 	  Stop 42, Cambridge, MA 02138, USA 
	 \and Argelander-Institut f\"ur Astromomie, Universit\"at Bonn, Auf dem 
	 	  H\"ugel 71, 53121 Bonn, Germany}
\abstract
{Circumstellar disks are expected to form early in the process that leads to the 
  formation of a young star, during the collapse of the dense molecular cloud 
  core. It is currently not well understood at what stage of the collapse the 
  disk is formed or how it subsequently evolves.}
{We aim to identify whether an embedded Keplerian protoplanetary disk 
  resides in the L1489~IRS system. Given the amount of envelope material still 
  present, such a disk would respresent a very young example of a protoplanetary 
  disk.}  
{Using the Submillimeter Array (SMA) we have observed the HCO$^+$ $J=$ 3--2 line 
  with a resolution of about 1$''$. At this resolution a protoplanetary disk 
  with a radius of a few hundred AUs should be detectable, if present. Radiative 
  transfer tools are used to model the emission from both continuum and line 
  data.}
{We find that these data are consistent with theoretical models of a collapsing 
  envelope and Keplerian circumstellar disk. Models reproducing both the SED and 
  the interferometric continuum observations reveal that the disk is inclined 
  by 40$^\circ$ which is significantly different to the surrounding envelope 
  (74$^\circ$).}
{This misalignment of the angular momentum axes may be caused by a gradient within 
  the angular momentum in the parental cloud or if L1489~IRS is a binary system 
  rather than just a single star. In the latter case, future observations looking 
  for variability at sub-arcsecond scales may be able to constrain these 
  dynamical variations directly. However, if stars form from turbulent cores, the 
  accreting material will not have a constant angular momentum axis (although the 
  average is well defined and conserved) in which case it is more likely to have 
  a misalignment of the angular momentum axes of the disk and the envelope.}

\maketitle

\section{Introduction}\label{intro}
Circumstellar disks constitute an integral part in the formation of low-mass 
protostars, being a direct result of the collapse of rotating molecular cloud 
cores and the likely birthplace of future planetary systems. When a 
molecular cloud core collapses to form a low-mass star, its initial angular 
momentum causes material to pile up on scales of $\sim 100$~AU in a 
circumstellar disk~\citep{cassen1981,terebey1984,adams1987,adams1988}. Due to 
accretion processes as the young stellar object (YSO) evolves, the disk grows in 
size while the envelope dissipates~\citep[e.g.,][]{basu1998,yorke1999}. In the 
more evolved stages of young stellar objects such disks can be imaged directly 
at optical and near-infrared wavelengths \citep[e.g.,][]{burrows1996, 
padgett1999} and their chemical and dynamical properties can be derived from 
comparison to submillimeter spectroscopic observations~\citep[e.g.,][]{thi2001}. 
Still, this latter method is often not unique, but requires \emph{a priori} 
assumptions about the underlying disk structure. In the earlier embedded stages 
this method is even more complex, because the disk is still deeply embedded in 
cloud material so that any signature of rotation in the disk itself, for 
example, is smeared out by emission that originates in the still collapsing 
envelope. By going to higher resolution using interferometric observations, as 
well as observing high density gas tracers, more reliable detections of 
protoplanetary disks can be made as direct imaging is approached 
\citep{rodriguez1998,qi2004,jorgensen2005}. In this paper we present such 
high-angular resolution submillimeter wavelength observations from the 
Submillimeter Array (SMA) of the embedded YSO L1489~IRS. We show how these 
observations place strong constraints on its dynamical structure and 
evolutionary status through detailed modeling. 

L1489~IRS (IRAS 04016+2610) is an intriguing protostar in the Taurus star 
forming region ($d=140$ pc) classified as a Class I YSO according to the
classification scheme of~\citet{Lada1984}. It is still embedded in a large 
amount of envelope material in which a significant amount of both infall and 
rotation has been observed on scales ranging from a few tenths of an AU out to 
several thousands of AUs~\citep{hogerheijde2001, boogert2002}. This large degree 
of rotation makes this source an interesting case study for the evolution of 
angular momentum during the formation of low-mass stars, potentially linking the 
embedded (Class 0 and I) and revealed (Class II and III) stages.

In a recent study by~\citet{brinch2007} (referred to as Paper~I), a model of 
L1489~IRS was presented based on data from a large single-dish molecular line 
survey~\citep{jorgensen2004}. This study was motivated by the intriguing 
peculiarities seen in L1489~IRS, such as the unusual large size and shape of the 
circumstellar material. The aim was to constrain the structure of its 
larger-scale infalling envelope and to place it in the right context of the 
canonical picture of low-mass star formation (see for example several reviews 
in~\citeauthor{reipurth2007} 2007).

The model presented in Paper~I describes a flattened envelope with an 
inspiraling velocity field, parameterized by the stellar mass and the (constant) 
angle of the field lines with respect to the azimuthal direction. A spherical 
temperature profile was adopted and the model did not explicitly contain a disk. 
The mass of the envelope is 0.09 M$_\odot$, adopted from 
\citet{jorgensen2002}. The use of such a ``global'' model is sufficient when 
working with single-dish data, where the emission is dominated by the emission 
from the large-scale envelope. With this description it was possible to 
accurately reproduce all the observed single-dish lines. 

What the study of Paper~I could not address, is whether a rotationally dominated 
disk is present on scales of the order of hundreds of AUs, although the amount 
of rotation which is inferred by the single-dish observations certainly suggests 
that a disk should have formed. In addition, two specific puzzles remain about 
the structure of L1489~IRS inferred on basis of the model when compared to other 
studies in the literature. First, the best fit was obtained with a central mass 
of 1.35 M$_\odot$, which is a very high value given that the luminosity of the 
star has been determined to be only 3.7 L$_\odot$~\citep{kenyon1993a}. Second, 
it was found that the best fit inclination of the system was 74$^\circ$. This is 
in agreement with the result from~\citet{hogerheijde2001} where they showed 
that in order to reproduce the observed aspect ratio the inclination cannot be 
less than 60$^\circ$. However, in a recent study,~\citet{eisner2005}, 
modeled the spectral energy distribution (SED) of L1489~IRS and demonstrated 
that this required a significantly different systemic inclination of only 
36$^\circ$.

In this paper, we try to address these issues through arc second scale 
interferometric observations of the dense gas tracer HCO$^+$ $J=3-2$ from the 
Submillimeter Array. These observations provide information on the gas dynamics 
on scales $\sim$100 AU and reveal an embedded protoplanetary disk. We show 
how careful modeling of the full SED from near-infrared through millimeter 
wavelengths can place strong constraints on the geometry of such a disk.

The outline of this paper is as follows: Sect.~\ref{obs} describes the details 
of the observations and data reduction while Sect.~\ref{results} presents the 
data. Sect.~\ref{analysis} introduces our model in which we have now included a 
disk and we also show how this model can fit all available observations 
including our new SMA data. Finally, in Sect.~\ref{discus} and~\ref{summary} we 
discuss the implications of our model and summarize our results.

\section{Observations and data reduction}\label{obs}
Observations were carried out at the Submillimeter Array (SMA)\footnote{The 
  Submillimeter Array is a joint project between the Smithsonian Astrophysical 
  Observatory and the Academia Sinica Institute of Astronomy and Astrophysics 
  and is funded by the Smithsonian Institution and the Academia Sinica} located 
on Mauna Kea, Hawaii. Our data set consists of two separate tracks of 
measurements in different array configurations. The first track was obtained on 
December 11, 2005. This track was done in the compact configuration resulting in 
a spatial resolution of $\sim$ 2.5$''$ with projected baselines ranging between 
10 and 62 k$\lambda$. A second track was obtained on November 28, 2006, in the 
extended configuration with the resolution of about 1$''$ and baselines between 
18 and 200 k$\lambda$. The resolution of the two configurations corresponds to 
linear sizes of 350 AU and 140 AU respectively. The synthesized beam 
size of the combined track using uniform visibility weighting is 0.9$'' \times$ 
0.7$''$ with a position angle of 78$^\circ$.

In both tracks the receiver was tuned to HCO$^+$ $J=$ 3--2 at 267.56 GHz. We 
used a correlator configuration with high spectral resolution across the line, 
providing a channel width of 0.2~kms$^{-1}$ over 0.104~GHz. The remainder of the 
2~GHz bandwidth of the SMA correlator was used to measure the continuum. No 
other lines in this band are expected to contaminate the continuum. We did not 
encounter any technical problems during observations and the weather conditions 
were excellent during both tracks. For the compact configuration track, 
$\tau_{225}$ was 0.06 and for the extended track $\tau_{225}$ was at 0.08.

Mars was used as flux calibrator in the first track, and Uranus for the second 
track. The quasars 3c454.3, 3c273, and 3c279 were used for passband calibration 
for both tracks, while the complex gains were calibrated using the two quasars 
3c111 and 3c84 located within 18~degrees from L1489~IRS. The calibrators
were measured every 20 minutes throughout both tracks. Their fluxes have been 
determined to be 3.1 and 2.6 Jy for 3c111 and 3c84 in the compact track, and 2.3 
and 2.2 Jy respectively in the extended track. For the gain calibration, a time 
smoothing scale of 0.7 hours was used which ensures that the large scale 
variations in the phase during the track are corrected for. The data do not
show any significant small scale variations (i.e., rapid fluctuations in the
phases or amplitudes). The quasars appear as point sources even at the longest 
baselines, which means that phase decorrelation due to atmospheric turbulence is 
negligible. The signal-to-noise of the calibrators is $>$50 per integration. 

The data were reduced using the MIR software package 
\citep{qi2005}\footnote{Available at 
http://cfa-www.harvard.edu/~cqi/mircook.html.}. Due to the excellent 
weather conditions, the data quality is very high and the data reduction 
procedure went smooth and unproblematic. All post-processing of the data, 
including combining the two visibility sets was done using the MIRIAD 
package~\citep{sault1995}\footnote{Available at 
http://bima.astro.umd.edu/miriad/miriad.html}. Relevant numbers are presented 
in Table~\ref{observations}.

In this paper we also make use of continuum measurements of L1489~IRS found in 
the literature ranging from the near-infrared~\citep{kenyon1993a, eisner2005, 
  padgett1999, park2002, whitney1997, myers1987, Kessler-Silacci2005} to 
  (sub)millimeter wavelengths~\citep{hogerheijde2000, Moriarty-Schieven1994, 
  motte2001, hogerheijde1997, ohashi1996, saito2001, lucas2000}.
\begin{table}
  \caption{Summary of the SMA observations. }\label{observations}
	\begin{tabular}{l c}
	\hline \hline
	Source					& 	L1489~IRS \\
	$\alpha,\delta$(2000)$^a$	& 	04:04:42.85, +26:18:56.3 \\
	Frequency\qquad\qquad	&	267.55762 GHz (1.12 mm) \\
	\hline
	Integrated line intensity		& 41.85 Jy beam$^{-1}$ km s$^{-1}$\\
	Peak line intensity				& 2.04 K\\
	\qquad Noise level (rms)		& 0.35 Jy beam$^{-1}$ \\
	Continuum flux					& 36.0$\pm$2.3 mJy    \\
	\qquad Noise level (rms) 		& 3.7 mJy beam$^{-1}$ \\
	Synthesized beam size			& $0.9''\times 0.7''$ \\
	\quad (Uniform weighting)       & \\
    \hline
  \end{tabular}

$^a$ Coordinates are given at the position where the continuum emission peaks.
\end{table}

\section{Results}\label{results}
The two tracks of observations provide us with two sets of $(u,v)$--points 
which, when combined and Fourier transformed, samples the image $(x,y)$ plane 
well from scales of 1 to 30$''$ (140 to 4200~AU). Emission on spatial 
scales greater than this is  filtered out by the interferometer due to its 
finite shortest spacing. This filtering is accounted for when comparing models to
the observations.

\begin{figure}
  \includegraphics[width=8.5cm]{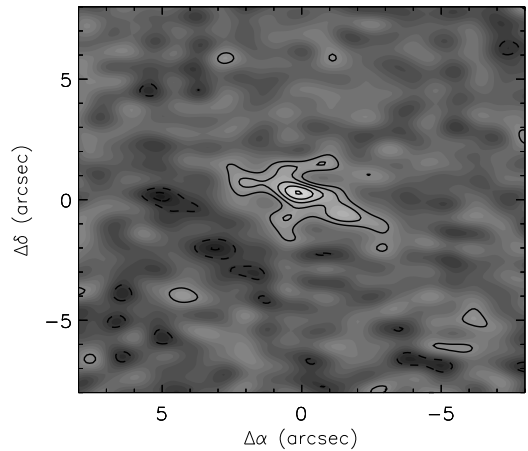}  
  \caption{Reconstructed image of the continuum emission at the highest 
  resolution (0.9$'' \times$ 0.7$''$). The contours are linearly spaced with 1$\sigma$ starting at 
  2$\sigma$ and dashed lines are negative contour levels.}\label{continuum}
\end{figure}

The measured continuum emission is shown as an image in Fig.~\ref{continuum}.
Previous attempts to measure the continuum at 1.1 mm with the BIMA 
interferometer were not successful~\citep{hogerheijde2001}. The SMA however,  
reveals a complex and detailed, slightly elongated, structure.

\begin{figure}
  \includegraphics[width=8.5cm]{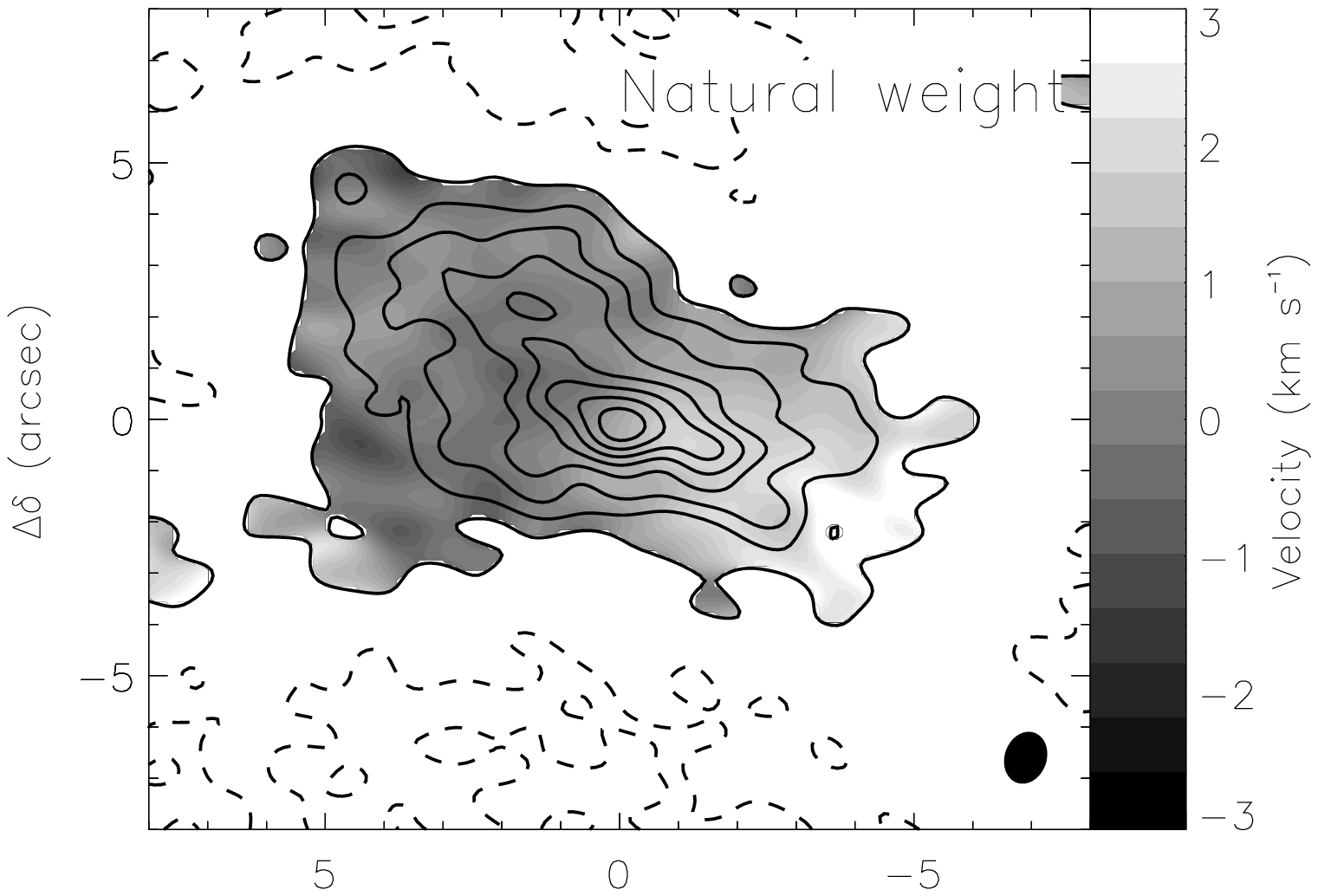}  
  \includegraphics[width=8.5cm]{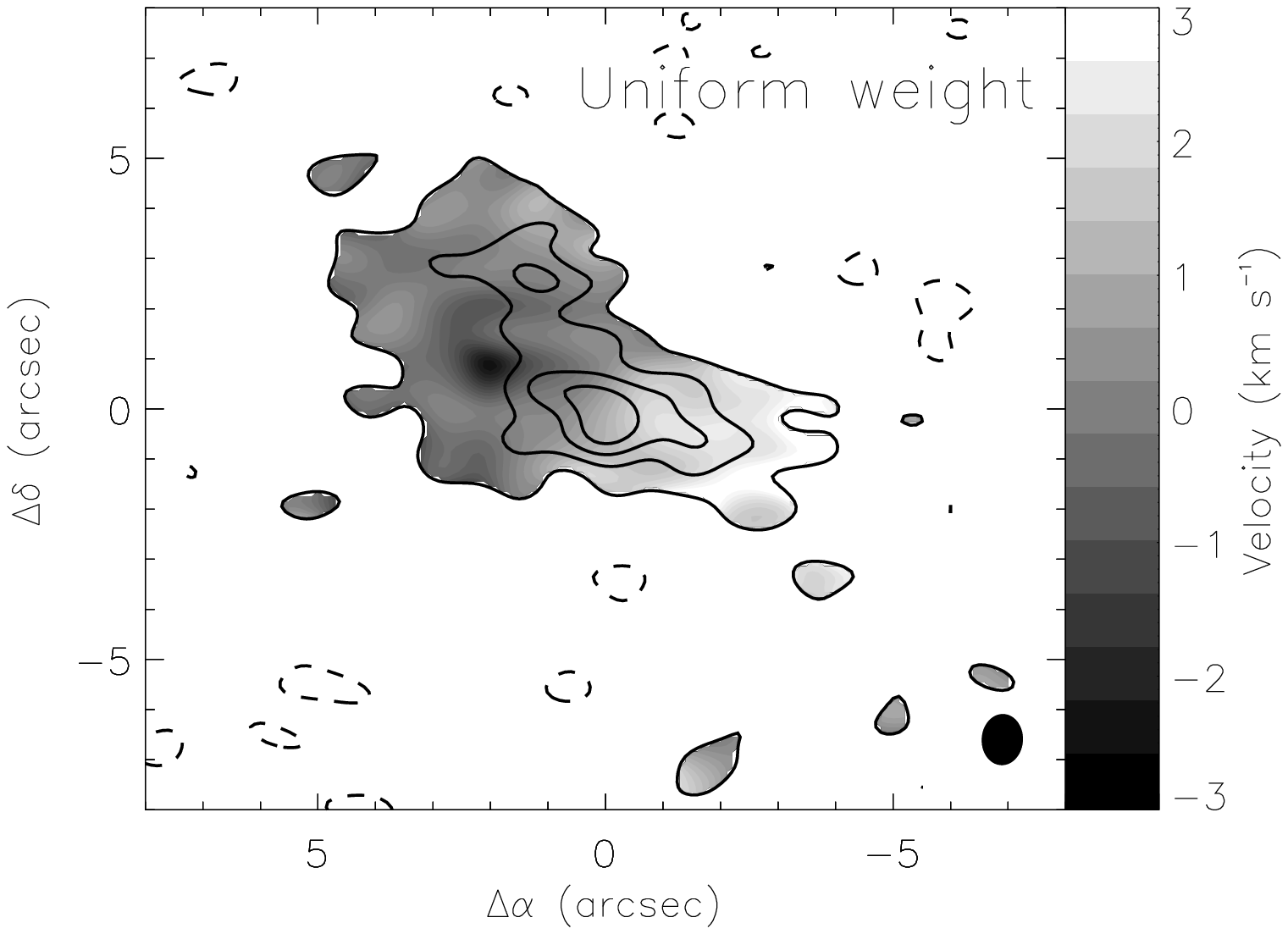}  
  \caption{Zero and first moment plots of the HCO$^+$ $J =$ 3--2 emission toward
  L1489~IRS. Contours are linearly spaced by 2$\sigma$ with a clip level of 
  2$\sigma$. $\sigma$ equals 1.1 and 2.0 Jy bm$^{-1}$ km s$^{-1}$ for 
  the natural and uniform weighting schemes respectively. Negative contours 
  appear as dashed lines.}\label{mom}
\end{figure}

Two reconstructed images of the HCO$^+$ emission are shown in Fig.~\ref{mom} 
using the \emph{natural} and \emph{uniform} weighting schemes, optimizing the 
signal-to-noise and angular resolution, respectively. In this figure, the zero 
moment map is plotted as solid contour lines and the first moment is shown as 
shaded contours. It is clear that both images reveal an elongated, flat 
structure. In the uniformly weighted image there is a large amount of asymmetry 
in the structure, which is less prominent in the naturally weighted image.

The velocity contours in the lower panel of Fig.~\ref{mom} are seen to be closed 
around a point which is offset with some 3$''$ from the peak of the emission.  
There is also clearly a gradient in the velocity field along the major axis of 
the object. This feature was previously reported by~\citet{hogerheijde2001} using 
interferometric observations of HCO$^+$ $J=$ 1--0, and low S/N HCO$^+$ $J=$ 3--2 
observations, from the BIMA and OVRO arrays. The gradient in the velocity field 
coincides almost perfectly with the long axis of the structure. When compared to 
the BIMA HCO$^+$ $J=$ 3--2 observations, the SMA data give 5--10 times better 
resolution. Furthermore, the BIMA data had to be self-calibrated using the 
HCO$^+$ $J=$ 1--0 image as a model and thus the resulting image was somewhat 
dependent of the structure of the lower excitation emission. The image we obtain 
from the SMA data is of considerably better quality.

\begin{figure}  
  \includegraphics[width=8.5cm]{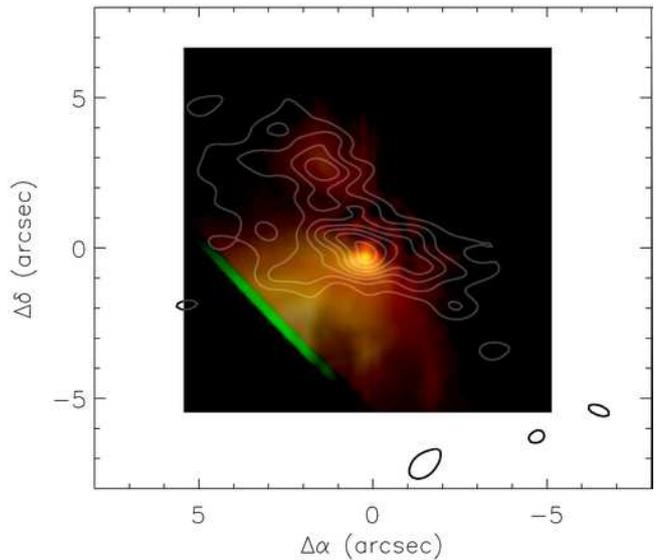}
  \caption{HCO$^+$ emission (1$\sigma$) contours superposed on the near-infrared 
  scattered light image~\citep{padgett1999}.}\label{image}
\end{figure}

In Fig.~\ref{image}, the zero moment emission contours have been superposed on the 
near-infrared scattered light image taken by the Hubble Space Telescope 
\citep{padgett1999}. The figure shows that many of the details in the SMA image
coincide with features seen in the scattered light image.

\begin{figure}
  \includegraphics[width=8.5cm]{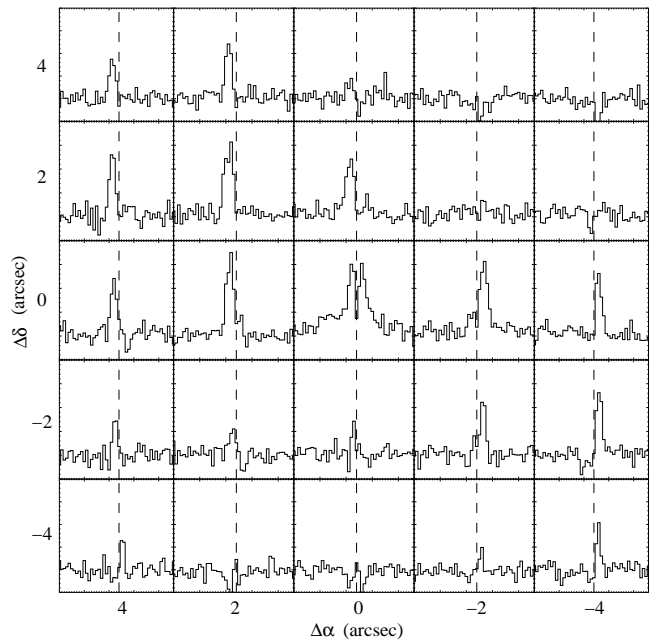}
  \caption{Spectral map of L1489~IRS. The spectra are evenly spaced with 2$''$.
           The bandwidth in each panel is 20 kms$^{-1}$. The scale on the 
		   $y$-axis goes from 0 to 2 Jybm$^{-1}$.}\label{specmap}
\end{figure}

The spectral map in Fig.~\ref{specmap} shows single spectra at positions offset 
by 2$''$ from each other. Each position shown here is the spectrum 
contained within one synthesized beam. The center most spectrum coincides with 
the peak of the emission in Fig.~\ref{mom}. Here is seen a very broad, double 
peaked line. Moving outward from the center, the lines become single peaked and 
offset in velocity space with respect to the systemic velocity. Perpendicular to 
the long axis of the structure, the line intensity falls off quickly. The lines 
shown in Fig.~\ref{specmap} are reconstructed using the natural weighting 
scheme. For the remainder of this paper we will use the uniformly weighted maps 
where the resolution is optimal.

\begin{figure}
  \includegraphics[width=8.5cm]{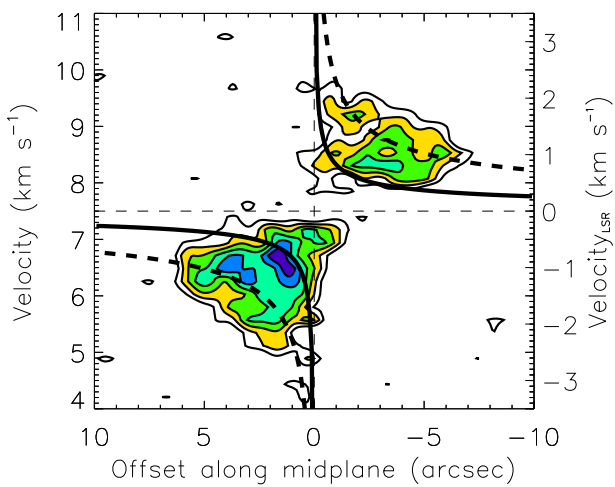}
  \caption{Position-Velocity (PV) diagram. This plot shows that the emission on the scales 
  measured by the SMA is entirely dominated by rotation. The black curves show 
  the Keplerian velocity, calculated from the dynamic mass (1.35 M$_\odot$) 
  obtained in Paper~I. The curves have been inclination corrected with 
  74$^\circ$ for the full lines and 40$^\circ$ for the dashed lines. 
  Contours start at 2$\sigma$ and 
  increase with 1$\sigma$ = 0.12 Jy beam$^{-1}$.}\label{pv}
\end{figure}

Fig.~\ref{pv} shows the emission in a position-velocity diagram, where the image 
cube has been sliced along the major axis to produce the intensity distribution 
along the velocity axis. The PV--diagram shows almost no emission in the second 
and fourth quadrant which is a strong indicator of rotation. The small amount of 
low velocity emission seen close to the center in the fourth quadrant may be 
accounted for by infalling gas. 

\section{Analysis}\label{analysis}
With the resolution provided by the SMA it is possible to probe the central 
parts of L1489~IRS on scales of $\sim$100 AU, where the protoplanetary disk is 
expected to be present (see Section~\ref{intro}). Furthermore, the 
HCO$^+$ $J=$ 3--2 transition traces H$_2$ densities of $10^6$~cm$^{-3}$ or 
more~\citep{schoier2005}, which are expected in the inner envelope and disk. The 
model of Paper~I was not explicitly made to mimic a disk, but rather to describe 
the morphology of the images on large scales.  Using this model, which works 
well on a global scale, we can explore how well it fits the SMA spectra when 
extrapolated to scales unconstrained by the single-dish observations. 

\begin{figure*}
  \begin{center}
  \includegraphics[width=16cm]{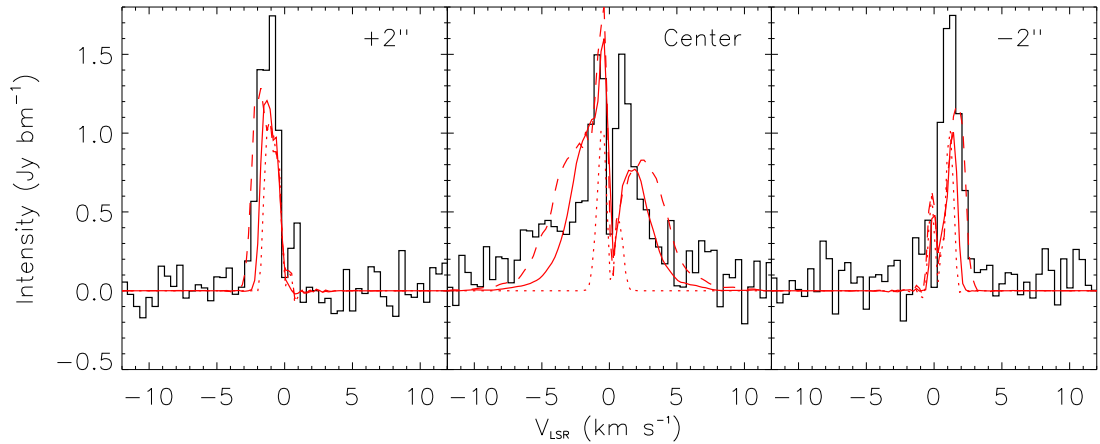}
  \caption{This figure shows three spectra from the SMA observations with model 
  spectra superposed. The two off-positions are chosen in the direction along 
  the long axis of the object. The offsets are chosen to be a resolution 
  element. Three models are also shown: The dashed line shows the unmodified 
  model from Paper~I, the full line is the model from Paper~I, but inclined at
  40$^\circ$, and the dotted line shows a model where HCO$^+$ is absent on 
  scales smaller than 200 AU.}\label{sma_specs}
\end{center}
\end{figure*}

Fig.~\ref{sma_specs} compares the SMA observations to the predictions of this 
model (dashed line). For doing this comparison, the model is 
imaged using the $(u,v)$-spacings from the observations, so that directly 
comparable spectra are obtained. While the spectra away from the center are 
fairly well reproduced in terms of line width, the center position is too wide, 
i.e., the model produces velocities that are too extreme compared to the 
measured velocities.

To check that this emission does in fact originate from scales less than 200 AU, 
a model where HCO$^+$ is completely absent within a radius of 200 AU is also 
shown in Fig.~\ref{sma_specs} with the dotted line. In this model the wide wings 
disappear and only a very narrow line is left. The two off-positions, which lie 
outside of the radius of 200~AU, are not affected minimally introducing this 
cavity.

Fig.~\ref{sma_specs} also shows a spectrum which is made from the Paper~I model, 
but inclined at 40$^\circ$ (full line). This is a considerably better fit than 
the other two models (again the two off-positions are little affected).  While 
the observations of the larger scale structure (Paper~I, 
\citet{hogerheijde2001}) demonstrates that the inclination of the flattened, 
collapsing envelope is $\sim$74$^\circ$, the SMA observations suggest that a 
change in the inclination occurs on scales of 100--200 AU, reflecting a 
dynamically different component there. On the other hand, since the model of
Paper~I was tuned to match the single-dish observation on scales of $\sim$1000 
AU or more and did not explicitly take the disk into account it is not 
unexpected that it does not fully reproduce the SMA observations.

\subsection{Introducing a disk model}\label{diskmod}
To reproduce also the interferometric observations we have modified the model 
from Paper~I on scales corresponding to the innermost envelope and disk: The 
improvements consists largely of two things, namely a different parameterization 
of the density distribution and the explicit inclusion of a disk.

The description used in Paper~I has a discontinuity for small values of $r$ and 
$\theta$: again, this is not a problem when working with the large scale 
emission, but becomes a problem when interfaced with a disk where a continuous 
transition from envelope to disk is preferred. The new parameterization we use 
for the envelope is adopted from~\citet{ulrich1976}. In this representation the 
envelope density is given by,
\begin{eqnarray}
\rho_{env}(r,\theta) &=& \rho_0 \left(\frac{r}{R_{rot}}\right)^{-1.5}
\left(1+\frac{\cos\theta}{\cos\theta_0}\right)^{-1/2} \nonumber   \\
&& \times \left( \frac{\cos\theta}{2 \cos\theta_0} +  \frac{R_{rot}}{r} \cos2
\theta_0\right)^{-1},
\end{eqnarray}
where $\theta_0$ is the solution of the parabolic motion of an infalling 
particle given by $r/R_{rot} (\cos \theta_0 -\cos \theta) / (\cos \theta_0 
\sin2 \theta_0)=1 $, $R_{rot}=150$ AU is the centrifugal radius of the envelope, 
and $\rho_0$ is the density on the equatorial plane and at the centrifugal 
radius. Since the outer radius, total mass, aspect ratio, and peak density are 
held fixed, the new parameterization is qualitatively similar to the Plummer-like
profile used in Paper~I. The numerical value of $R_{rot}$ is chosen so
that the maximum deviation in density in the ($r,\theta$)-plane only reaches 
20\% in the most dense parts (outside of the disk cavity) and has an average 
deviation of less than 5\%.

\begin{figure}
  \begin{center}
  \includegraphics[width=8.5cm]{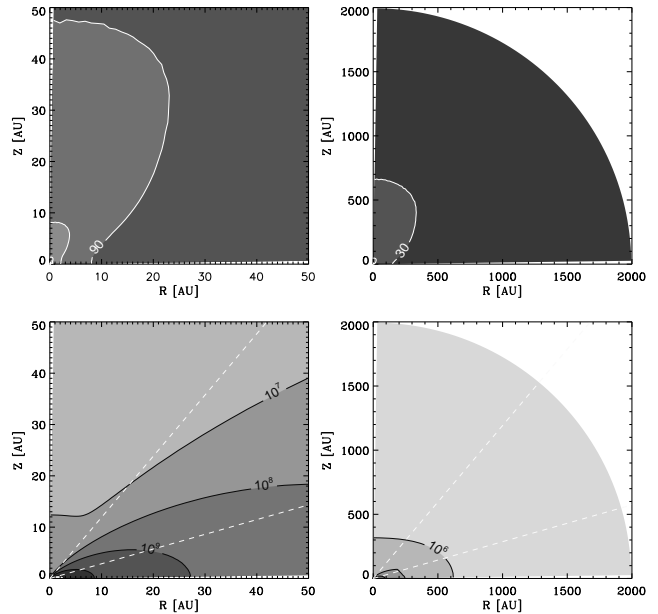}
  \caption{A slice through the model of L1489~IRS: The top panels show the
  temperature on different scales and the bottom panels show the corresponding
  density profiles.}\label{diskstruct}
\end{center}
\end{figure}

In addition to this envelope, a generic disk is also introduced in the model. 
The density of the disk is given by,
\begin{eqnarray}
\rho_{disk}(r,\theta)=\frac{\Sigma_0 \, (r/R_0)^{-1}}{\sqrt{2\pi} \,
H_0}  \exp \left\{-\frac{1}{2} \left[ \frac{r \, \cos \theta}{H_0}
\right]^2 \right\}, 
\end{eqnarray}
where $\theta$ is the angle from the axis of symmetry, $R_0$ is the disk outer 
radius and $H_0$ is the scale height of the disk. The model thus has four free 
parameters: disk radius, pressure scale height, mass (taken as a fraction of 
the total mass of the circumstellar material), and inclination. In order to 
maximize the mid-infrared flux, we use a flat disk, i.e., with no increase in 
scale height with radius and fix the scale height to 0.25. The outer radius is 
fixed to 200~AU, the distance from the center of the emission in Fig.~\ref{mom} 
to the position of the closed velocity contour, thereby effectively reducing 
the number of free parameters to two. This outer disk radius is not a
strongly constrained parameter since neither the SED nor the model spectra are 
influenced much by changes in this value. It is only possible to rule out the 
extreme cases, where the disk either gets so small that it is no longer 
influencing the SED or where it gets so big that it holds a significant fraction 
of the total mass. We find that the best fit is provided by a disk mass of 
$M_{disk}=4\times 10^{-3} $~M$_\odot$. A cross section of the model is shown in 
Fig.~\ref{diskstruct}. 

The velocity field in the envelope is similar to the one used in Paper~I, where
the velocity field was parameterized in terms of a central mass and an angle 
between the velocity vector and the azimuthal direction, so the the ratio of 
infall to rotation could be controlled by adjusting this angle. The best fit 
parameter values obtained in Paper~I of 1.35M$_\odot$ for the central mass and 
flow lines inclined with 15$^\circ$ with respect to the azimuthal direction, 
are used. No radial motions are allowed in the disk: only full Keplerian motion 
is present in the region of the model occupied by the disk.

In order to produce synthetic observations which can be directly compared to the 
data, we use numerical radiative transfer tools. The 3D continuum radiative 
transfer code RADMC~\citep{dullemond2004} is used to calculate the scattering 
function and the temperature structure in the analytical axisymmetric density 
distribution. In these calculations, the luminosity measurement 
by~\citet{kenyon1993a} of 3.7 L$_\odot$ is used for the central source.
The solution is then ``ray--traced'' using RADICAL \citep{dullemond2000} to 
produce the spectral energy distribution and the continuum maps at the observed 
frequencies. In both RADMC and RADICAL we assume certain properties of 
the dust. We use the dust opacities that give the best fit to extinction 
measurements in dense cores (Pontoppidan et al., in prep). 

The same density and temperature structure are afterward given as input to the 
excitation and line radiative transfer code RATRAN~\citep{hogerheijde2000a} 
which is used to calculate the spatial and frequency dependent HCO$^+$ $J=$ 
3--2 emission. The HCO$^+$ models are post-processed with the MIRIAD tasks 
\emph{uvmodel, invert, clean,} and \emph{restore} to simulate the actual SMA 
observations.

\subsection{Modeling the continuum emission}\label{sec:sed}
\begin{figure}
  \begin{center}
  \includegraphics[width=8.5cm]{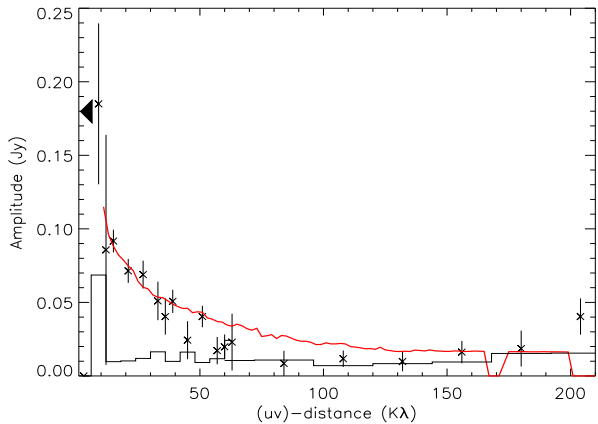}
  \caption{The averaged $(u,v)$--amplitudes of the continuum at 1 mm in both 
  	   	   compact and extended configurations. The full line show our 
		   model. The compact configuration visibilities cover the 
		   $(u,v)$--distance up to 60 k$\lambda$. The black triangle marks the 
		   total flux at 1 mm~\citep{Moriarty-Schieven1994}. The histogram
		   indicates the zero-signal expectation values.}\label{dustamps}
\end{center}
\end{figure}

In Fig.~\ref{dustamps}, we compare our model to the continuum observations. We 
calculate $(u,v)$--amplitudes at 1.1 mm which we plot on top of the observed 
amplitudes. The result is in good agreement with the data, 
suggesting that the dust emission is well-described by our parameterization down 
to scales of $\sim$~100~AU. This comparison is not very dependent on the 
inclination, since the continuum emission is quite compact. For this particular 
figure, an inclination of 40$^\circ$ is assumed.

Recently,~\citet{eisner2005} modeled the SEDs of a number of Class I objects in 
Taurus including L1489~IRS. The model used in their work is parameterized 
differently from ours, but it essentially describes a similar structure. As 
pointed out above, \citeauthor{eisner2005} find a best fit inclination 
of 36$^\circ$ in contrast to the inclination of 74$^\circ$ found in Paper~I.

\begin{figure}
  \begin{center}
  \includegraphics[width=8.5cm]{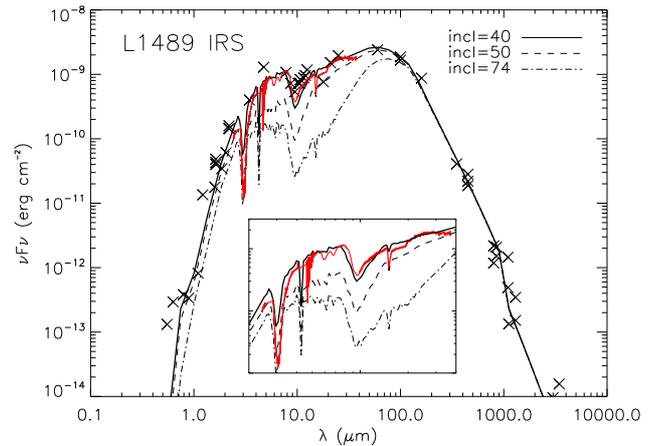}
  \caption{The model spectral energy distribution assuming
    inclinations of 40$^\circ$ (full line) and 50$^\circ$ and
    74$^\circ$ (broken lines) plotted on top of flux measurements from
    the literature (marked by crosses) and the Spitzer/IRS spectrum
    from 2 -- 40 microns (red line).}\label{sed}
\end{center}
\end{figure}

To test the result of~\citet{eisner2005} we calculated the SED using the 
described model with only the system inclination as a free parameter 
(Fig.~\ref{sed}). The best fit is found for an inclination of about 40$^\circ$, 
in good agreement with the result of~\citet{eisner2005}. The models with
inclinations of 50$^\circ$ and 74$^\circ$, are also plotted in
Fig.~\ref{sed} in order to show how the SED depends on inclination.

Note that the quality of our best fit to the SED is comparable to the fit 
presented by~\citet{eisner2005}, whereas the fit becomes rapidly worse with 
increasing inclination larger than 40$^\circ$, thus bringing about disagreement 
with the results from~\citep{hogerheijde2001} and Paper~I. The three different 
models plotted in Fig.~\ref{sed} cannot be distinguished for wavelength above
$\sim$60~$\mu$m, which corresponds to a temperature of about 40~K 
using Wiens displacement law. This temperature occurs on radial distances of 
approximately 100 AU from the central object which means that outside of this 
radius, the SED is no longer sensitive to the inclination. We interpret this 
behavior as a change in the angular momentum axis on disk scales, causing the
disk to be inclined with respect to the envelope.

\subsection{Modeling the HCO$^+$ emission}\label{sec:hcop}
\begin{figure}
  \begin{center}
  \includegraphics[width=8.5cm]{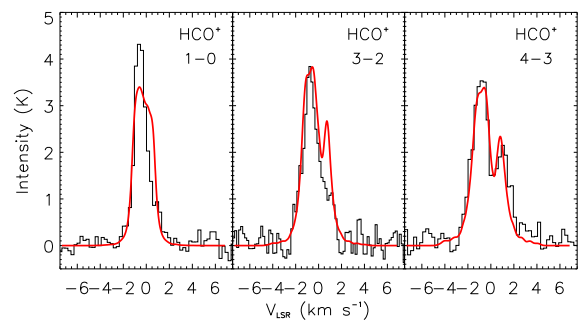}
  \caption{Three transitions of HCO$^+$ observed by single-dish telescope (see
  	  	   Paper~I for details) with our model superposed. The quality of the
		   fit is comparable to the fit in Paper~I although a disk model has now 
		   been included.}\label{specs}
\end{center}
\end{figure}

In Paper~I it was found that the characteristic double peak feature of the 
HCO$^+$ $J=$ 4--3 line could not be reproduced with an inclination below 
70$^\circ$. On the other hand including a disk inclined by 40$^\circ$ into the 
model of Paper~I does not alter the fit to the single dish lines 
(Fig~\ref{specs}): the geometry and the velocity field of the material at 
scales smaller than $\sim$300 AU does not influence the shape of these lines. 
This fit is not perfect though as it still overestimates the red-shifted wing in 
the $J=$ 3--2 line and the width of the $J=$ 1--0 line slightly, but the quality
of the fit is similar to that presented in Paper~I.

\begin{figure}
  \begin{center}
  \includegraphics[width=8.5cm]{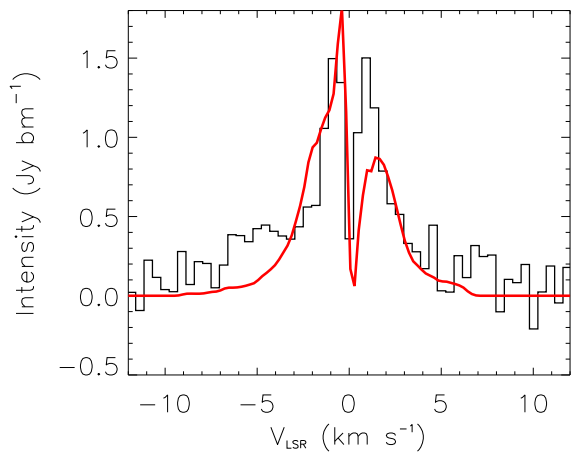}
  \caption{The SMA spectrum towards the center of L1489~IRS with our model,
  consisting of a disk which is inclined with respect to the envelope, 
  superposed. As in Fig.~\ref{sma_specs}, the model has been imaged with 
  the $(u,v)$-spacings from the observations.}\label{bestfit}
\end{center}
\end{figure}

We need to test how well this tilted disk model works with the line observations 
from the SMA. It was shown in Sect.~\ref{analysis} that the off-position spectra 
are weakly dependent on changes in the inclination and that the same is true when 
using the model where the disk is tilted with respect to the envelope. The 
central position however, is seen in Fig.~\ref{bestfit} to be very well 
reproduced in terms of line width and wing shape by this model. The model 
spectrum is slightly more asymmetric than the data, which means that the
infall to rotation ratio is not quite correct on disk scales. The magnitude of 
the velocity field projected onto our line of sight is correct though, since the 
width of the line is well fit.

\begin{figure}
  \begin{center}
  \includegraphics[width=8.5cm]{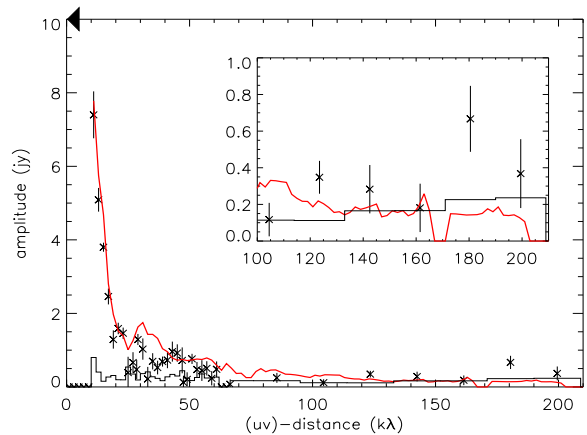}
  \caption{Similar to Fig.~\ref{dustamps} but for the HCO$^+$ $J=$
    3--2 measurements. The triangle marks the total flux obtained from
    the single dish observations. The amplitudes have been averaged
    over $\pm 5$~kms$^{-1}$ relative to the systemic velocity. The 
	histogram indicates the zero-signal expectation values. The insert is an 
	enlargement of the longest baselines.}\label{uvamps}
\end{center}
\end{figure}

We also compared the averaged $(u,v)$--amplitudes to evaluate the quality of our 
model similar to the procedure for the continuum. Fitting the 
$(u,v)$--amplitudes in this way tests whether the model produces the right amount 
of emission at every scale. The best fit is shown in Fig.~\ref{uvamps}. The 
zero-spacing flux has been shown in this plot as well, marked by the black 
triangle. This method has previously been used to study the abundance variations 
of given molecules in protostellar envelopes, in particular imaging directly
where significant freeze-out occurs in protostellar envelopes
\citep{jorgensen2004a,jorgensen2005a}. The model does a good job reproducing 
the observed HCO$^+$ brightness distribution on almost all scales with a 
constant abundance, except around 20--40 k$\lambda$ where it overestimates the 
amount of observed flux. This correspond to a radius of about 1000 AU, that is 
well outside of the disk. On the other hand single-dish observations of low-mass 
protostellar envelopes suggest that these are scales where significant 
freeze-out in particular of CO may occur at temperatures $\lesssim 20-30$~K, 
which because of the gas-phase chemical relation between CO and HCO$^+$ also
reflects directly in the observed distribution of HCO$^+$
\citep[e.g.,][]{jorgensen2004,jorgensen2005}. Despite this, the difference between 
the model prediction and observations is small with this constant abundance, 
suggesting that the amount of freeze-out is small in L1489~IRS. This is in agreement 
with the single-dish studies of \cite{jorgensen2002,jorgensen2004,jorgensen2005}, 
who generally found little depletion in the envelopes of L1489~IRS and other 
Class~I sources in contrast to the more deeply embedded, Class~0, protostars.

The model parameter values obtained in this section including values from 
Paper~I are summarized in Table~\ref{parameters}. The best fit
values that are derived in this paper are tuned by hand rather than 
systematically optimized by a $\chi^2$ minimization method. Therefore we can   
only give an estimate on the uncertainties. However, Fig.~\ref{sed}
shows that the SED fit gets rapidly worse when changing the inclination, and 
given the strong dependence of the envelope inclination on the single-dish 
lines, we estimate that both angles are accurate to within 10$^\circ$. For the
disk mass and radius, the uncertainties on the parameter values are less well 
defined for reasons given in Section~\ref{diskmod}, but we estimate an accuracy
within a factor of 2 in each of these parameters.

\begin{table}
  \caption{Model parameter values}\label{parameters}
	\begin{tabular}{l c}
	\hline \hline
    Envelope & Value \\
	\hline
	Outer radius$^a$ 				& 2000 AU\\
	Envelope mass$^a$				& 0.093 M$_\odot$\\
	Inclination$^{a,b}$ 			& 74$^\circ$\\
	HCO$^+$ abundance$^a$\qquad 	& 3.0$\times$10$^{-9}$\\
	\quad & \quad\\
	\hline
	Disk & Value \\
	\hline
	Radius$^b$ 						& 200 AU\\	
	Disk mass$^b$ 					& 4$\times$10$^{-3}$ M$_\odot$\\
	Inclination$^b$ 				& 40$^\circ$\\
	Scale height$^c$ 				& z=0.25R\\
	Central mass$^a$ 				& 1.35 M$_\odot$\\
	Central luminosity$^c$			& 3.7 L$_\odot$\\
	Distance$^c$ 					& 140 pc\\
	\hline
  \end{tabular}

$^a$ Values from Paper I\\
$^b$ Fitted values\\
$^c$ Values adopted from the literature\\
\end{table}

\section{Discussion}\label{discus}
\begin{figure}
  \begin{center}
  \includegraphics[width=7.5cm]{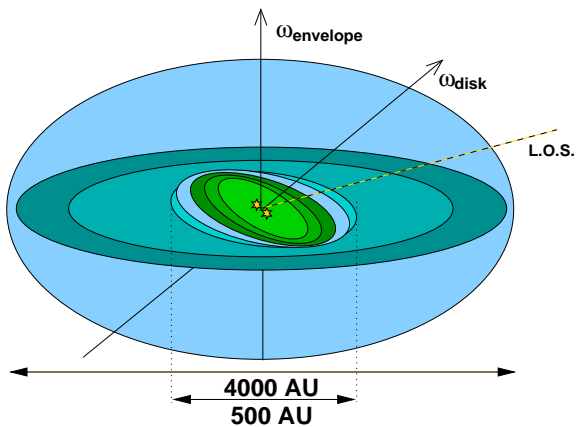}
  \caption{An (exaggerated) illustration of the proposed model where the angular 
  momentum axis is changing with radius. The line of sight is illustrated by the
  dashed line. }\label{cartoon}
\end{center}
\end{figure}

It seems that depending on the physical size scales that we probe, the solution 
favors a different inclination. When taking \emph{all available} observations 
into account we need to introduce a model where the angular momentum vector of 
the disk is misaligned with the angular momentum vector of the (rotationally 
flattened) envelope. We adopt a two component model (illustrated in 
Fig.~\ref{cartoon}), disk and envelope, but in reality, the angular momentum 
axis may indeed be a continuous function of radius on the scales of the disk 
($\lesssim$ a few hundred AU). The origin of such a misalignment can actually be
explained in a simple way by considering the initial conditions of the collapse.

The state of a pre-stellar core before the gravitational collapse begins is 
typically described by a static sphere with a solid body rotation perturbation
\citep{terebey1984}. Due to strict angular momentum conservation, such a model 
would be perfectly aligned throughout the collapse and consequently disk 
formation models are often described numerically by axisymmetric representations
\citep[e.g.,][]{yorke1999}.

However, there is no \emph{a priori} reason why a pre-stellar core should rotate 
as a solid body. Of course the cloud has an average angular momentum which on a 
global scale determines the axis of rotation. The cloud does not collapse 
instantly to form a disk though, but rather from the inside and out as described 
by~\citet{shu1977}, and therefore a shell of material from deep within the 
cloud, which may very well have an average angular momentum that is different 
from the global average, will collapse and form a disk before material from 
further out has had a change to accrete yet. Actually, if the parental cloud is 
turbulent, it is to be expected that the accreting material has randomly 
oriented angular momentum and a misalignment of the angular momentum on different 
scales is more likely than a perfect alignment. In this way, a system similar to 
the model proposed to describe L1489~IRS in this paper can be formed.

To obtain a gradient in the angular momentum as initial condition for the 
collapse, one may consider a cloud which is not spherically symmetric or one which 
does not collapse around its geometrical center but rather around an over dense 
clump offset from the center. The former scenario may indeed be true for the 
case of L1489~IRS which is seen to be dynamically connected to a neighboring 
cloud (which was also modeled in Paper~I to explain the excess of cold emission 
seen in some of the low-$J$ lines). Actually, the uniformly weighted moment map
(Fig.~\ref{mom}) shows significant asymmetry with a secondary emission peak to 
the north-east.

It is interesting to note that the HCO$^+$ emission from the SMA observations 
agree very well with the near-infrared image (Fig.~\ref{image}): the secondary 
peak, a few arc seconds northeast of the main continuum and HCO$^+$ peaks, nicely 
coincides with a bright spot in the scattered light image and also the shape of 
the cavity towards the south follows the contours of the HCO$^+$ emission. The
same details are not revealed in the naturally weighted SMA image, in which 
shorter baselines are given more weight. We thus conclude that by going to the 
extended SMA configuration, it is possible to probe structure in YSOs on the same 
scales as can be resolved by large near-infrared telescopes such as the Hubble 
Space Telescope. The structure seen in both images, however, also emphasizes 
that for a fully self-consistent description of L1489~IRS, a global
non-axisymmetric model has to be considered.

Another (non-exclusive) explanation of a misaligned disk would be that L1489~IRS 
formed as a triple stellar system and, that due to gravitational interaction, 
one of the stars was ejected. L1489~IRS would thus be a binary system as 
suggested by~\citet{hogerheijde2000}. The loss of angular momentum due to the 
ejection would result in a rearrangement of the remaining binary, which could 
``drag'' the inner viscous disk along. Such a scenario has been investigated 
numerically by~\citet{larwood1996}. In the case of L1489~IRS, it would have to 
be a very close binary with a separation of no more than a few AUs since the 
near-infrared scattered light image~\citep{padgett1999} with a resolution of 
$\sim$0.2$''$ (30 AU), does not reveal multiple sources. The ratio 
of binary separation to disk radius is therefore significantly lower than the 
cases investigated by~\citet{larwood1996} and it is therefore not clear whether 
similar effects could be present in L1489~IRS.

If indeed the system is binary, it would resolve the issue that we find 
a central mass of 1.35 M$_\odot$ while the luminosity is estimated to be 3.7 
L$_\odot$~\citep{kenyon1993a}. A single young star that massive would require a 
much higher luminosity, but two stars of 0.6--0.7 M$_\odot$ would fit nicely 
with the estimated luminosity since luminosity is not a linear function of mass 
for YSOs.

To test whether L1489~IRS indeed is a binary and to constrain the innermost disk 
geometry on AU scales, measurements of temporal variability in super high 
resolution are needed. The timescale for variations of a possible binary is of 
the order of years depending on the exact separation of the stars (assuming 
scales of 1 AU). Such observations would, for example, be feasible with ALMA.

\section{Conclusion}\label{summary}
In this paper, we have presented high angular resolution interferometric 
observations of the low-mass Class I YSO L1489~IRS. The observations reveal a 
rotationally dominated, very structured central region with a radius of about 
200--300 AU. We interpret this as a young Keplerian disk which is still deeply
embedded in envelope material. This conclusion is supported by a convincing fit 
with a disk model to the SED.

We conclude further that the inclination of the disk is not aligned with the 
inclination of the flattened envelope structure, due to the possibility that 
L1489~IRS is a binary system and/or that the average angular momentum axis of 
the cloud is not aligned with the angular momentum axis of the dense core that 
originally collapsed to form the star(s) plus the disk.

We find that a disk with a mass of $4\times 10^{-3}$ M$_\odot$, a radius of 200 
AU, and a pressure scale height of z=0.25R is consistent with both the SED and 
the HCO$^+$ observations. Only a small amount of chemical depletion of HCO$^+$ 
is allowed for, due to a slight over-estimate of the $(u,v)$--amplitudes at 
20--40 k$\lambda$ by our model, in agreement with the results from previous 
single-dish studies and the nature of L1489~IRS as a Class I YSO.

The combination of a detailed modeling of the SED with spatially resolved line 
observations, which contains information on the gas kinematics, appears to be a very 
efficient way of determining the properties of disks, especially embedded disks 
which are not directly observable.\\

\noindent \emph{Acknowledgments} The authors would like to thank Kees Dullemond 
for making his code available to us. CB is supported by the European Commission 
through the FP6 - Marie Curie Early Stage Researcher Training programme. JKJ 
acknowledges support from an SMA fellowship. AC was supported by a fellowship 
from the European Research Training Network ``The Origin of Planetary Systems'' 
(PLANETS, contract number HPRN-CT-2002-00308). The research of MRH and AC is 
supported through a VIDI grant from the Netherlands Organiztion for Scientific 
Research.\\

\bibliographystyle{aa}
\bibliography{/home/brinch/papers/references}

\begin{thebibliography}{43}
\expandafter\ifx\csname natexlab\endcsname\relax\def\natexlab#1{#1}\fi

\bibitem[{{Adams} {et~al.}(1987){Adams}, {Lada}, \& {Shu}}]{adams1987}
{Adams}, F.~C., {Lada}, C.~J., \& {Shu}, F.~H. 1987, \apj, 312, 788

\bibitem[{{Adams} {et~al.}(1988){Adams}, {Shu}, \& {Lada}}]{adams1988}
{Adams}, F.~C., {Shu}, F.~H., \& {Lada}, C.~J. 1988, \apj, 326, 865

\bibitem[{{Basu}(1998)}]{basu1998}
{Basu}, S. 1998, \apj, 509, 229

\bibitem[{{Boogert} {et~al.}(2002){Boogert}, {Hogerheijde}, \&
  {Blake}}]{boogert2002}
{Boogert}, A.~C.~A., {Hogerheijde}, M.~R., \& {Blake}, G.~A. 2002, \apj, 568,
  761

\bibitem[{{Brinch} {et~al.}(2007){Brinch}, {Crapsi}, {Hogerheijde}, \&
  {J{\o}rgensen}}]{brinch2007}
{Brinch}, C., {Crapsi}, A., {Hogerheijde}, M.~R., \& {J{\o}rgensen}, J.~K.
  2007, \aap, 461, 1037

\bibitem[{{Burrows} {et~al.}(1996){Burrows}, {Stapelfeldt}, {Watson}, {Krist},
  {Ballester}, {Clarke}, {Crisp}, {Gallagher}, {Griffiths}, {Hester},
  {Hoessel}, {Holtzman}, {Mould}, {Scowen}, {Trauger}, \&
  {Westphal}}]{burrows1996}
{Burrows}, C.~J., {Stapelfeldt}, K.~R., {Watson}, A.~M., {et~al.} 1996, \apj,
  473, 437

\bibitem[{{Cassen} \& {Moosman}(1981)}]{cassen1981}
{Cassen}, P. \& {Moosman}, A. 1981, Icarus, 48, 353

\bibitem[{{Dullemond} \& {Dominik}(2004)}]{dullemond2004}
{Dullemond}, C.~P. \& {Dominik}, C. 2004, \aap, 417, 159

\bibitem[{{Dullemond} \& {Turolla}(2000)}]{dullemond2000}
{Dullemond}, C.~P. \& {Turolla}, R. 2000, \aap, 360, 1187

\bibitem[{{Eisner} {et~al.}(2005){Eisner}, {Hillenbrand}, {Carpenter}, \&
  {Wolf}}]{eisner2005}
{Eisner}, J.~A., {Hillenbrand}, L.~A., {Carpenter}, J.~M., \& {Wolf}, S. 2005,
  \apj, 635, 396

\bibitem[{{Hogerheijde}(2001)}]{hogerheijde2001}
{Hogerheijde}, M.~R. 2001, \apj, 553, 618

\bibitem[{{Hogerheijde} \& {Sandell}(2000)}]{hogerheijde2000}
{Hogerheijde}, M.~R. \& {Sandell}, G. 2000, \apj, 534, 880

\bibitem[{{Hogerheijde} \& {van der Tak}(2000)}]{hogerheijde2000a}
{Hogerheijde}, M.~R. \& {van der Tak}, F.~F.~S. 2000, \aap, 362, 697

\bibitem[{{Hogerheijde} {et~al.}(1997){Hogerheijde}, {van Dishoeck}, {Blake},
  \& {van Langevelde}}]{hogerheijde1997}
{Hogerheijde}, M.~R., {van Dishoeck}, E.~F., {Blake}, G.~A., \& {van
  Langevelde}, H.~J. 1997, \apj, 489, 293

\bibitem[{{J{\o}rgensen}(2004)}]{jorgensen2004a}
{J{\o}rgensen}, J.~K. 2004, \aap, 424, 589

\bibitem[{{J{\o}rgensen} {et~al.}(2005{\natexlab{a}}){J{\o}rgensen}, {Bourke},
  {Myers}, {Sch{\"o}ier}, {van Dishoeck}, \& {Wilner}}]{jorgensen2005a}
{J{\o}rgensen}, J.~K., {Bourke}, T.~L., {Myers}, P.~C., {et~al.}
  2005{\natexlab{a}}, \apj, 632, 973

\bibitem[{{J{\o}rgensen} {et~al.}(2002){J{\o}rgensen}, {Sch{\" o}ier}, \& {van
  Dishoeck}}]{jorgensen2002}
{J{\o}rgensen}, J.~K., {Sch{\" o}ier}, F.~L., \& {van Dishoeck}, E.~F. 2002,
  \aap, 389, 908

\bibitem[{{J{\o}rgensen} {et~al.}(2004){J{\o}rgensen}, {Sch{\" o}ier}, \& {van
  Dishoeck}}]{jorgensen2004}
{J{\o}rgensen}, J.~K., {Sch{\" o}ier}, F.~L., \& {van Dishoeck}, E.~F. 2004,
  \aap, 416, 603

\bibitem[{{J{\o}rgensen} {et~al.}(2005{\natexlab{b}}){J{\o}rgensen},
  {Sch{\"o}ier}, \& {van Dishoeck}}]{jorgensen2005}
{J{\o}rgensen}, J.~K., {Sch{\"o}ier}, F.~L., \& {van Dishoeck}, E.~F.
  2005{\natexlab{b}}, \aap, 435, 177

\bibitem[{{Kenyon} {et~al.}(1993){Kenyon}, {Calvet}, \&
  {Hartmann}}]{kenyon1993a}
{Kenyon}, S.~J., {Calvet}, N., \& {Hartmann}, L. 1993, \apj, 414, 676

\bibitem[{{Kessler-Silacci} {et~al.}(2005){Kessler-Silacci}, {Hillenbrand},
  {Blake}, \& {Meyer}}]{Kessler-Silacci2005}
{Kessler-Silacci}, J.~E., {Hillenbrand}, L.~A., {Blake}, G.~A., \& {Meyer},
  M.~R. 2005, \apj, 622, 404

\bibitem[{{Lada} \& {Wilking}(1984)}]{Lada1984}
{Lada}, C.~J. \& {Wilking}, B.~A. 1984, \apj, 287, 610

\bibitem[{{Larwood} {et~al.}(1996){Larwood}, {Nelson}, {Papaloizou}, \&
  {Terquem}}]{larwood1996}
{Larwood}, J.~D., {Nelson}, R.~P., {Papaloizou}, J.~C.~B., \& {Terquem}, C.
  1996, \mnras, 282, 597

\bibitem[{{Lucas} {et~al.}(2000){Lucas}, {Blundell}, \& {Roche}}]{lucas2000}
{Lucas}, P.~W., {Blundell}, K.~M., \& {Roche}, P.~F. 2000, \mnras, 318, 526

\bibitem[{{Moriarty-Schieven} {et~al.}(1994){Moriarty-Schieven}, {Wannier},
  {Keene}, \& {Tamura}}]{Moriarty-Schieven1994}
{Moriarty-Schieven}, G.~H., {Wannier}, P.~G., {Keene}, J., \& {Tamura}, M.
  1994, \apj, 436, 800

\bibitem[{{Motte} \& {Andr{\'e}}(2001)}]{motte2001}
{Motte}, F. \& {Andr{\'e}}, P. 2001, \aap, 365, 440

\bibitem[{{Myers} {et~al.}(1987){Myers}, {Fuller}, {Mathieu}, {Beichman},
  {Benson}, {Schild}, \& {Emerson}}]{myers1987}
{Myers}, P.~C., {Fuller}, G.~A., {Mathieu}, R.~D., {et~al.} 1987, \apj, 319,
  340

\bibitem[{{Ohashi} {et~al.}(1996){Ohashi}, {Hayashi}, {Kawabe}, \&
  {Ishiguro}}]{ohashi1996}
{Ohashi}, N., {Hayashi}, M., {Kawabe}, R., \& {Ishiguro}, M. 1996, \apj, 466,
  317

\bibitem[{{Padgett} {et~al.}(1999){Padgett}, {Brandner}, {Stapelfeldt},
  {Strom}, {Terebey}, \& {Koerner}}]{padgett1999}
{Padgett}, D.~L., {Brandner}, W., {Stapelfeldt}, K.~R., {et~al.} 1999, \aj,
  117, 1490

\bibitem[{{Park} \& {Kenyon}(2002)}]{park2002}
{Park}, S. \& {Kenyon}, S.~J. 2002, \aj, 123, 3370

\bibitem[{{Qi}(2005)}]{qi2005}
{Qi}, C. 2005, The MIR Cookbook, The Submillimeter Array/Harvard-Smithsonian
  Center for Astrophysics

\bibitem[{{Qi} {et~al.}(2004){Qi}, {Ho}, {Wilner}, {Takakuwa}, {Hirano},
  {Ohashi}, {Bourke}, {Zhang}, {Blake}, {Hogerheijde}, {Saito}, {Choi}, \&
  {Yang}}]{qi2004}
{Qi}, C., {Ho}, P.~T.~P., {Wilner}, D.~J., {et~al.} 2004, \apjl, 616, L11

\bibitem[{Reipurth {et~al.}(2007)Reipurth, Jewitt, \& Keil}]{reipurth2007}
Reipurth, B., Jewitt, D., \& Keil, K., eds. 2007, Protostars and Planets V
  (University of Arizona Press)

\bibitem[{{Rodr{\'{\i}}guez} {et~al.}(1998){Rodr{\'{\i}}guez}, {D'Alessio},
  {Wilner}, {Ho}, {Torrelles}, {Curiel}, {G{\'o}mez}, {Lizano}, {Pedlar},
  {Cant{\'o}}, \& {Raga}}]{rodriguez1998}
{Rodr{\'{\i}}guez}, L.~F., {D'Alessio}, P., {Wilner}, D.~J., {et~al.} 1998,
  \nat, 395, 355

\bibitem[{{Saito} {et~al.}(2001){Saito}, {Kawabe}, {Kitamura}, \&
  {Sunada}}]{saito2001}
{Saito}, M., {Kawabe}, R., {Kitamura}, Y., \& {Sunada}, K. 2001, \apj, 547, 840

\bibitem[{{Sault} {et~al.}(1995){Sault}, {Teuben}, \& {Wright}}]{sault1995}
{Sault}, R.~J., {Teuben}, P.~J., \& {Wright}, M.~C.~H. 1995, Astronomical Data
  Analysis Software and Systems IV, 77, 433

\bibitem[{{Sch{\"o}ier} {et~al.}(2005){Sch{\"o}ier}, {van der Tak}, {van
  Dishoeck}, \& {Black}}]{schoier2005}
{Sch{\"o}ier}, F.~L., {van der Tak}, F.~F.~S., {van Dishoeck}, E.~F., \&
  {Black}, J.~H. 2005, \aap, 432, 369

\bibitem[{{Shu}(1977)}]{shu1977}
{Shu}, F.~H. 1977, \apj, 214, 488

\bibitem[{{Terebey} {et~al.}(1984){Terebey}, {Shu}, \& {Cassen}}]{terebey1984}
{Terebey}, S., {Shu}, F.~H., \& {Cassen}, P. 1984, \apj, 286, 529

\bibitem[{{Thi} {et~al.}(2001){Thi}, {van Dishoeck}, {Blake}, {van Zadelhoff},
  {Horn}, {Becklin}, {Mannings}, {Sargent}, {van den Ancker}, {Natta}, \&
  {Kessler}}]{thi2001}
{Thi}, W.~F., {van Dishoeck}, E.~F., {Blake}, G.~A., {et~al.} 2001, \apj, 561,
  1074

\bibitem[{{Ulrich}(1976)}]{ulrich1976}
{Ulrich}, R.~K. 1976, \apj, 210, 377

\bibitem[{{Whitney} {et~al.}(1997){Whitney}, {Kenyon}, \&
  {Gomez}}]{whitney1997}
{Whitney}, B.~A., {Kenyon}, S.~J., \& {Gomez}, M. 1997, \apj, 485, 703

\bibitem[{{Yorke} \& {Bodenheimer}(1999)}]{yorke1999}
{Yorke}, H.~W. \& {Bodenheimer}, P. 1999, \apj, 525, 330

\end{thebibliography}
\end{document}